# Cooperative Dynamics of an Artificial Stochastic Resonant System

Yasushi Hotta[*], Teruo Kanki, Naoki Asakawa, Hitoshi Tabata[1], and Tomoji. Kawai

*Institute of Scientific and Industrial Research, Osaka University, 8-1 Mihogaoka, Ibaraki, Osaka 567-0047, Japan*

[1]*Department of Bioengineering, School of Engineering, University of Tokyo, 7-3-1 Hongo, Bunkyoku, Tokyo 113-8656, Japan*

We have investigated cooperative dynamics of an artificial stochastic resonant system, which is a recurrent ring connection of neuron-like signal transducers (NST) based on stochastic resonance (SR), using electronic circuit experiments. The ring showed quasi-periodic, tunable oscillation driven by only noise. An oscillation coherently amplified by noise demonstrated that SR may lead to unusual oscillation features. Furthermore, we found that the ring showed synchronized oscillation in a chain network composed of multiple rings. Our results suggest that basic functions (oscillation and synchronization) that may be used in the central pattern generator of biological system are induced by collective integration of the NST element.

*E-mail address: hotta@sanken.osaka-u.ac.jp

Stochastic resonance (SR) is a phenomenon in which noise enhances the response of a nonlinear system to a weak signal.[1-3] It has been identified in various physical systems such as electronic circuits [2, 4] and natural systems such as neurons.[1, 2, 5-10] Noise is intuitively interfering factor, but is can be used to provide a functional benefit of signal coherency amplification in the stochastic resonant systems. Recent studies suggest the potential technological application of SR, such as for sensors to detect weak signal,[10] image reproduction in visual perception,[11, 12] and nonlinear memory storages,[13, 14] and aim for the common utilization of this phenomenon in the engineering field.[14] Although it is important to develop SR for common use, a general scheme for making functional and integrated systems based on SR has yet to be investigated. In order to broaden the use of SR, we explore an integration scheme of SR elements.

As informative examples, biological systems may functionally use this phenomenon in neurophysiological systems such as sensory receptors in crayfish,[6] crickets,[7] and humans,[8] and behaviors such as paddlefish feeding on zooplankton and human balance control.[15, 16] Given that the neuron is an SR element,[1, 2, 5-10] nervous systems could be considered one integrated system comprising SR elements. Neurons connect within functionally correlative groups that further connect within groups of groups, and represent an integral element of large-scale functional systems such as neural networks. In this study, inspired by observations of SR in biological systems, we propose a scheme to integrate a neuron-like signal transducer (NST) with SR into a recurrent ring system, which is a simplified version of a recurrent neural network.[17] We demonstrate the integration using an electronic circuit and investigate cooperative dynamics of the recurrent ring. Furthermore, the recurrent rings were investigated for behavior in a chain connection in order to confirm the capacity of more integrated systems based on the ring.

Figures 1(a) and 1(b) show a circuit diagram of the NST and its characteristics. Our electronic circuits were fabricated using standard linear operational amplifier (OPA: TL082CP, Texas Instruments). The comparator consists of a Schmitt trigger, a bistable hysteretic device with dual threshold of $V_{th}$ and $-V_{th}$. It produces a constant low- ($V_L$) or high-state ($V_H$) voltage as output signals depending on the input signal level. If the input signal crosses the threshold ($V_{th}$ or $-V_{th}$) of the comparator [Fig. 1(b), (i)], the comparator state is switched between $V_H$ and $V_L$, resulting in a binary signal train for a time series [Fig. 1(b), (ii)]. The output signal is decayed to the ground (GND) level by the derivation circuit with a time constant, indicating a temporal decay of the effective output level similar to that of a neuron [Fig. 1(b), (iii)]. The reference signals [subthreshold modulations, Gaussian white noise (GWN) and trigger signals] in this experiment were provided by function generators (Wave Factory 1946, NF Corporation). Time-series data from the simulation circuits were recorded using a computer-controlled analogue-to-digital converter (USB-6008, National Instruments). For the NST, we characterized SR by the signal-to-noise ratio (SNR) of output signals as a function of the amplitude of GWN. In order to calculate SNR values, the signal train from the circuits was recorded for noise with various amplitudes and processed by fast Fourier transform to yield a power spectrum. As shown in Fig. 1(c), the SNR vs noise amplitude relationship produced a bell-shaped curve, realizing an SR characteristic.

Using this NST element, we made a recurrent ring connection composed of four elements directionally connected with a coupling constant $w_{cc}$ in ring geometry [Fig.

2(a)]. $w_{cc}$ is defined as $V_H/V_{th}$ (or $V_L/-V_{th}$, where $V_L = -V_H$). The same value of $w_{cc} = 2$ (fixed as $V_H = 200$ mV and $V_{th} = 100$ mV) was used for all the SR elements in this experiment, unless otherwise specified.

We report that quasi-periodic oscillation resulted in the ring circuit by only adding GWN. The NST does not have an oscillation function, and the oscillation property arises from the geometric effect of the recurrent ring connection. In other words, the ring can rectify noise to a periodic signal, like a Brownian motor.[18, 19] This point is quite different from previously reported studies of the stochastic oscillator in computer simulations, which used the oscillator as a base element.[20] Figures 2(b) and 2(c) show time-series data of the ring output and power spectrum as a parameter of the GWN amplitude. We can see periodic patterns and power spectrum peaks. The frequency of the oscillation increased with increasing noise amplitude [Fig. 2(d)]. This correlation indicates that noise contributes significantly to the oscillation. In contrast, the coherency of the oscillation was amplified at a noise amplitude of 1.2 V. SNR of coherency of the output [Fig. 2(e)] represented the bell-shaped pattern for noise amplitude, indicating that the ring was still in the regime of SR. The increase in oscillation frequency and decrease in coherency were also observed when $w_{cc}$ was reduced with fixing of the noise amplitude. It is thought that large amplitude of noise reduces the effective coupling constant between the two NSTs.

The ring was further characterized for response to an external trigger signal comprising a sine wave with various frequencies. Figure 3(a) shows typical time-series data of the ring response at different frequencies. The ring circuit showed burst-like behavior at 0.1 Hz [Fig. 3(a), (i)]. With an increase in frequency, the oscillation began to tune to the external signal with higher-order synchronization, and then it was tuned up to several dozen Hz as shown in Fig. 3(a), (ii) (1Hz) and (iii) (3 Hz). The oscillation became irregular over 100 Hz [Fig. 3(a), (iv), 400 Hz]. In order to measure synchronization, we calculated the cross-correlation $R_{xy}$ between the trigger signal and the ring output, as a function of the frequency of the trigger signal. In Fig. 3(b), the $R_{xy}$ value was large for a frequency ranging from 2 Hz to a few hundred Hz. The ring was tunable to the trigger signal without changing any circuit parameter, and the maximum point was determined by the time constant of the differentiation circuit. The tunable flexibility served to extend the operating range of the oscillation frequency of the ring.

In order to investigate a more collective system, the rings were further characterized in the chain network [Fig. 4(a)]. Figure 4(b) shows oscillation properties of the ring in the chain for noise amplitudes of 0.9, 1.2, and 1.6 V. The rings showed cooperative oscillation following the pattern of the most upstream ring. During the signal propagation in the chain, the phase progressively advanced as the more downstream ring. This phase advance was significantly observed at 1.2 V [Fig. 4(b), (ii), but not at 0.9 or 1.6 V.

To interpret the phase advance in the regime of SR, we used as an example a model of a coupled multi-limb system controlled by oscillation of the chain system.[21] The phase-advanced synchronization produced cooperative motion of two connected limbs. The cooperative motion propagated to downstream limbs, realizing a smooth meandering motion for the whole of the multi-limb system, as shown in Fig. 4(c). The cooperative motion arising from the phase advance of the oscillators was enhanced by noise, and corresponds to the SR-type behavior.

Considering the circuit design, SR should have arisen only in NSTs. But it also arose in the output of the ring and the behavior of the chain. This means that a collective system composed of the base elements should have SR-type behavior in system functions if we introduce mechanisms leading to SR in the base element. This property of *succession* provides flexible features in system functions, which may arise from the robust nature of SR. In this study, such a feature appeared in the tunable oscillation of the ring. Our results are similar to phenomena observed in biological experiments, in which noise added to sensory receptors caused SR in the behavior of living individuals.[15, 16] The analogy implies that our procedure of integrating isolated elements mimics biological systems. We believe that this scheme to integrate NSTs is important in the utilization of accumulated function regarding SR.

Lastly, the scheme to utilize SR could contribute to an energy saving for devices. Recently, the electric power consumption density of large-scale integrated circuits has increased and is comparable to that of a cooking hot plate, and will reach the same level of the sun's surface in the near future.[22] A dominant factor is that the driving voltage of a transistor is less reducible to overcome system noise, even with an increase of the transistor element density on a tip. Since a noise-induced system is unnecessary to overcome noise, the driving voltage can be dramatically reduced at the element level. The scheme may therefore allow for a total saving of the energy consumption of a system.

**Acknowledgement** This research was supported by "Special Coordination Funds for Promoting Science and Technology: Yuragi Project" of the Ministry of Education, Culture, Sports, Science and Technology, Japan.

# REFERENCES

1) K. Wiesenfeld and F. Moss: Nature **373** (1995) 33.

2) L. Gammaitoni, P. Hänggi, P. Jung, and F. Marchesoni: Rev. Mod. Phys. **70** (1998) 223.

3) Note: Previous studies manifested minimal components to achieve SR in a nonlinear system, which are a form of threshold in a bistable system, a source of noise, and an input of subthreshold modulation (weak signal). For mechanism of the SR, the bistable potential is tilted up and down when subthreshold modulation is added to the system, thereby asymmetrically raising and lowering the potential threshold of one minimum state and the other, respectively. In addition, a non-zero level of noise stochastically assists the state transition over the threshold, and the response of the system to weak signal is enhanced. Since an appropriate level of noise raises the efficiency of the assistance, the efficiency defined by the signal-to-noise ratio as a function of noise amplitude shows a clear bell-shaped pattern.

4) S. Fauve and F. Heslot: Phys. Lett. **97**A (1983) 5.

5) A. Longtin, A. Bulsara, and F. Moss: Phys. Rev. Lett. **67** (1991) 656.

6) J. K. Douglass, L. Wilkens, E. Pantazelou, and F. Moss: Nature **365** (1993) 337.

7) J. E. Levin and J. P. Miller: Nature **380** (1996) 165.

8) J. J. Collins, T. T. Imhoff, and P. Grigg: Nature **383** (1996) 770.

9) J. J. Collins, T. T. Imhoff, and P. Grigg: J. Neurophysiol. **76** (1996) 642.

10) J. J. Collins, C. C. Chow, and T. T. Imhoff: Nature **376** (1995) 236.

11) E. Simonotto, M. Riani, C. Seife, M. Roberts, J. Twitty, and F. Moss: Phys. Rev. Lett. **78** (1997) 1186.

12) B. Kosko and S. Mitaim: Neural Networks **16** (2003) 755.

13) M. F. Carusela, R. P. J. Perazzo, and L. Romanelli: Phys. Rev. E **64** (2001) 031101.

14) R. L. Badzey and P. Mohanty: Nature **437** (2005) 995.

15) D. F. Russell, L. A. Wilkens, and F. Moss: Nature **402** (1999) 291.

16) A. Priplata, J. Niemi, M. Salen, J. Harry, L. A. Lipsitz, and J. J. Collins: Phys. Rev. Lett. **89** (2002) 238101.

17) N. Asakawa, Y. Hotta, T. Kanki, T. Kawai, and H. Tabata: Phys. Rev. E, in press.

18) R. D. Astumian: Sci. Am. July (2001) 57.

19) P. Reimann and P. Hänggi: Appl. Phys. A **75** (2002) 169.

20) S. K. Han, W. S. Kim, and H. Kook: Phys. Rev. E **58** (1998) 2325.

21) Note: It is known that animals have such organs of oscillator systems, composed of a neural network (called a central pattern generator) in their spinal cord [E. R. Kndel, J. H. Schwartz, and T. M. Jessell: *Principles of Neural Science* (McGraw-Hill, New York, 2000) P.737.]

22) P. P. Gelsinger: Proc. IEEE Int. Solid-State Circuits Conf., vol. XLIV 2001, P.22.

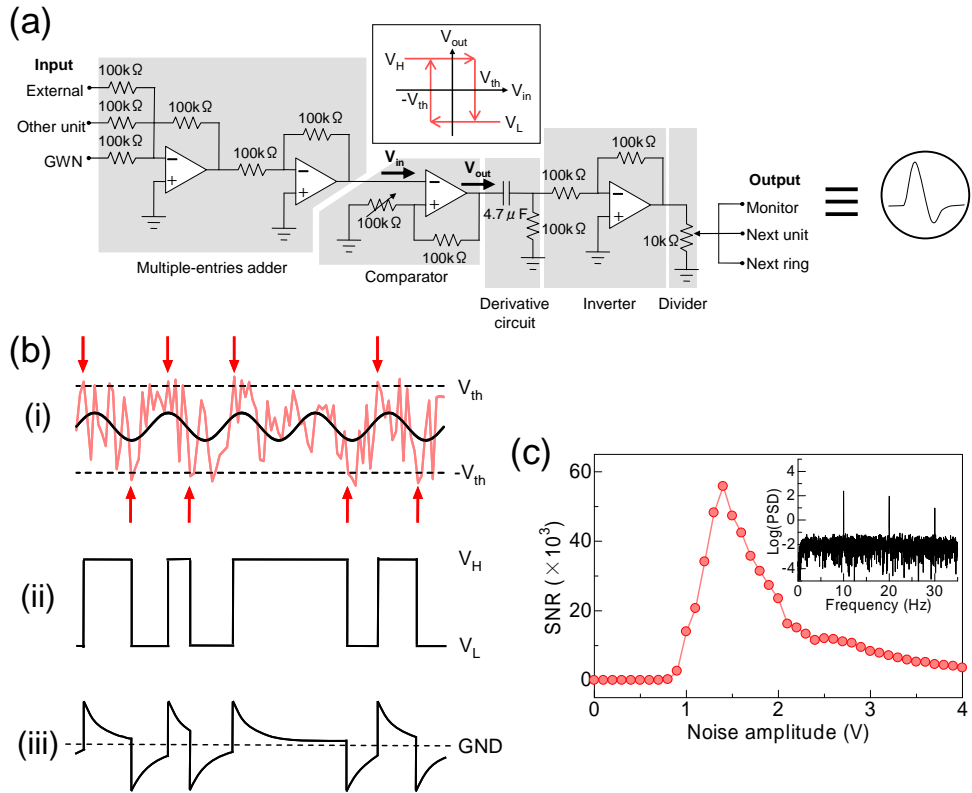

**FIG 1. (a)** Analog electric circuit diagram for the NST. The experimental circuit of the NST consists of a multiple-entries adder, a comparator, and a derivative circuit. The multiple-entries adder and comparator were composed of OPA circuits in the double-inverter and Schmitt trigger configurations, respectively. The inset shows a hysteresis characteristic of the comparator. The derivative circuit is fabricated using a capacitance-resistance circuit. After the derivative circuit, the signal was sent to one more inverter circuit in order to obtain a non-inverted signal against the input because the comparator works as an inverter. Two trimmer resistances were used in order to vary the threshold value (and hysteresis width) and $w_{cc}$: one was located between the feedback resistance and the ground in the Schmitt trigger circuit, and the other was used as a voltage divider after the inverter circuit. We symbolized the NST on the right side.

**(b)** (i) Subthreshold modulation (black) and Gaussian white noise, η(t), are summed at the adder, and the summed signal (red) is sent to the comparator. The red arrows indicate threshold-crossing points. (ii) If the summed signal crosses the threshold of the comparator, the output voltage of the comparator is stochastically switched between the low- ($V_L$) and high-state ($V_H$) voltage, resulting in a binary signal train as output. (iii) The binary signal train is differentiated at a derivative circuit to produce a neuron-like pulse.

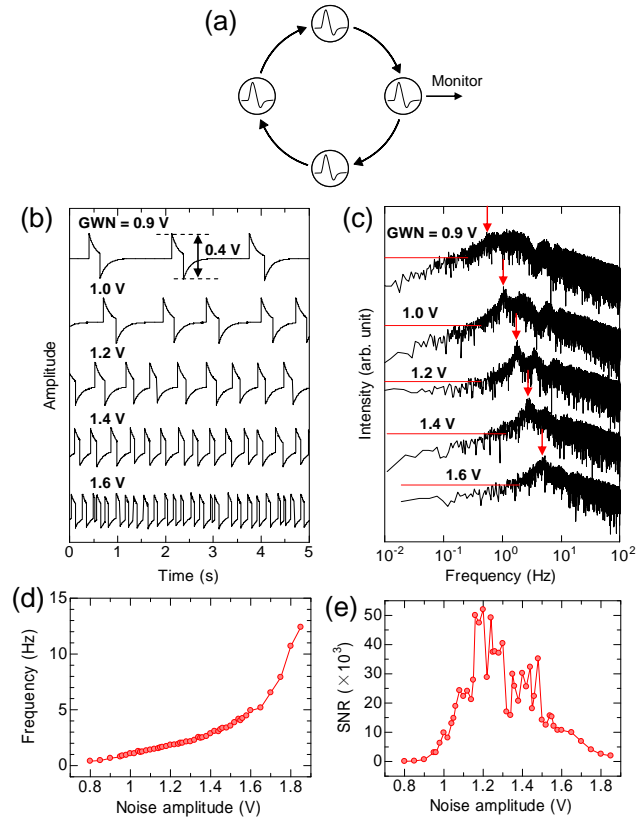

**FIG 2.** Oscillation properties of a ring.

**(a)** A recurrent ring connection of the NSTs.

**(b)** Typical time-series data of binary signal trains from the ring when GWN was only applied. The ring showed oscillation at approximately ±0.8 V above the GWN amplitude.

**(c)** Coherency of oscillations for various noise amplitudes was analyzed using the SNR value, which was obtained from power spectra normalized to restrict the time-series data to 256 oscillations. Red lines and arrows indicate background noise levels and each peak position of the oscillation patterns, respectively.

**(d)** and **(e)** Oscillation frequency and SNR for the oscillation signal of the ring as a function of noise amplitude, respectively.

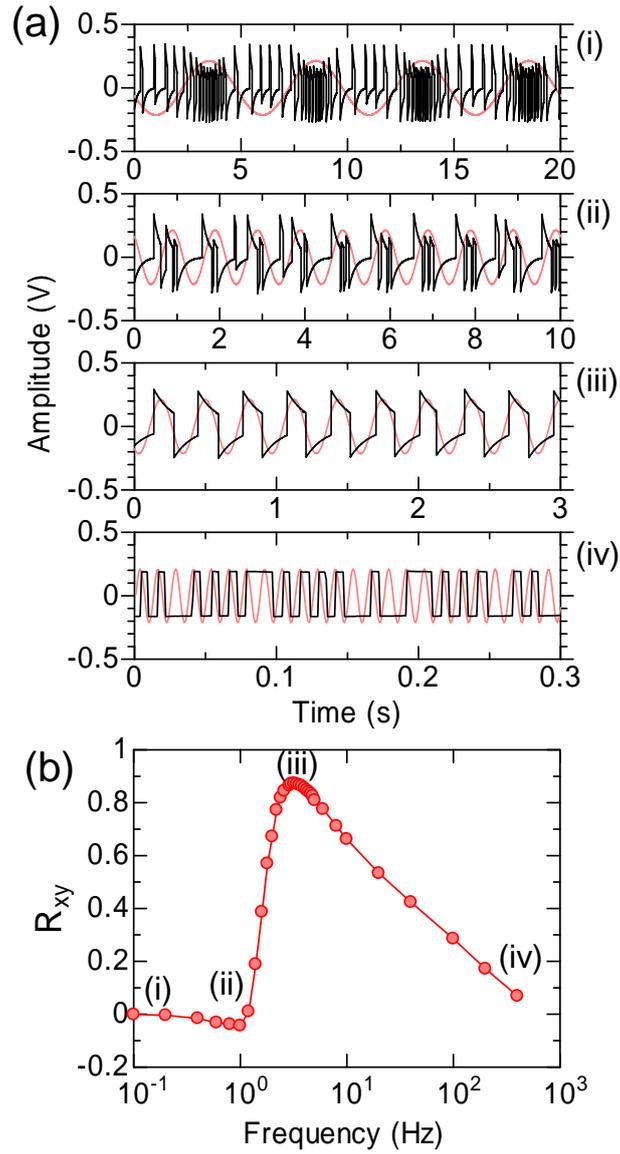

**FIG 3.** Tunable-oscillation behavior of the ring.

**(a)** The oscillation responses for an external signal with various frequencies. The noise amplitude is fixed at 1.2 V, corresponding to the maximum point of SNR in Fig. 2C.

**(b)** Cross-correlation between external and output signal trains as a function of external signal frequency. The labeled points (i) - (iv) correspond to the plots in **(a)**.

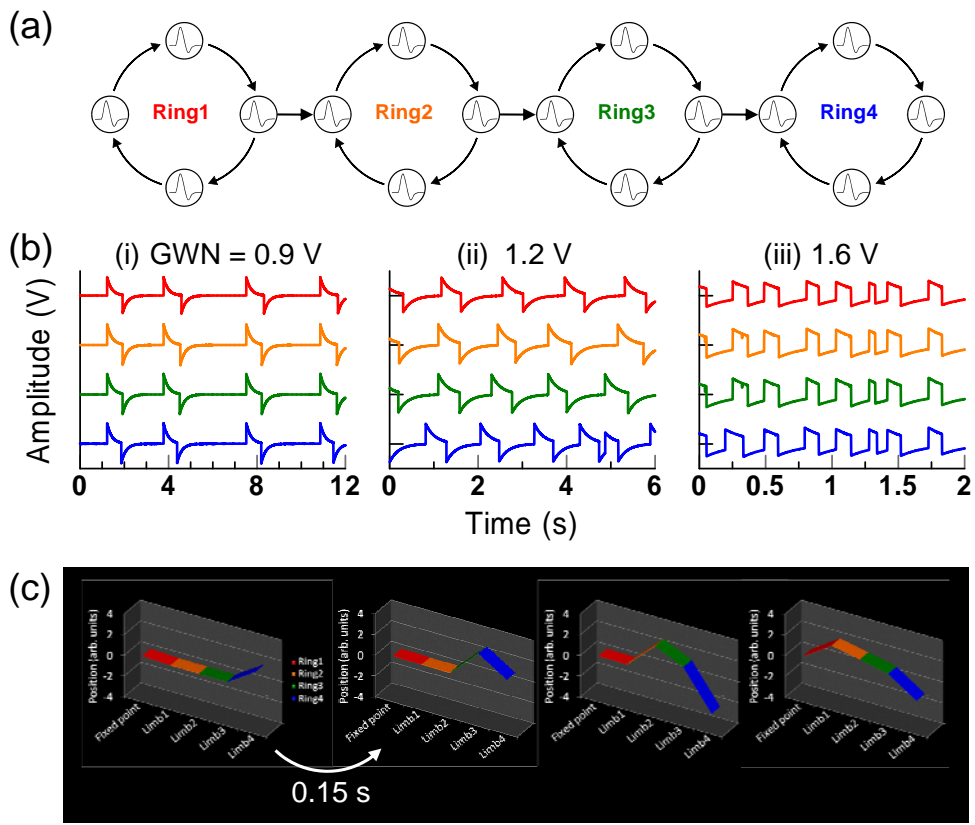

**FIG4.** Oscillation properties of the ring in a chain network.

**(a)** Configuration of a chain network of rings. The chain was composed of four rings by connecting one output of an NST in the upstream rings to one input in the downstream rings.

**(b)** Oscillation pattern of each ring for noise amplitudes of (i) 0.9, (ii) 1.2, and (iii) 1.6 V. Ring1, Ring2, Ring3, and Ring4 are represented by red, orange, green and blue lines, respectively.

**(c)** Demonstration of successive flatfish swimming motion, controlled using the ring oscillator chain. The interval between two frames is 0.15 s. The position along the vertical axis corresponds to the oscillation amplitude of the ring.

13